\documentclass[fleqn]{article}
\usepackage{graphicx} 
\usepackage{subfig}
\usepackage{authblk}
\usepackage{amsmath}
\usepackage[dvipsnames]{xcolor}
\usepackage{hyperref}
\usepackage[margin=1.2in]{geometry}
\usepackage{lineno}
\newcommand{\boldsize}[1]{
  {\noindent\fontsize{11}{11}\selectfont \textbf{#1}}
}
\title{Bridging chemistry and Gaussian boson sampling: A photonic hierarchy of approximations for molecular vibronic spectra}

\author[1]{Jan-Lucas Eickmann\thanks{Corresponding author, janlucas@mail.uni-paderborn.de}}
\author[1]{Kai-Hong Luo}
\author[1]{Mikhail Roiz}
\author[1]{Jonas Lammers}
\author[1]{Simone Atzeni}
\author[1]{Cheeranjiv Pandey}
\author[1]{Florian Lütkewitte}
\author[2]{Reza G. Shirazi}
\author[1]{Fabian Schlue}
\author[1]{Benjamin Brecht}
\author[2]{Vladimir V. Rybkin\thanks{Corresponding author, vladimir.rybkin@quantumsimulations.de}}
\author[1]{Michael Stefszky}
\author[1]{Christine Silberhorn}
\affil[1]{Paderborn University, Integrated Quantum Optics, Institute for Photonic Quantum Systems (PhoQS), Warburger Str. 100, 33098 Paderborn, Germany}
\affil[2]{HQS Quantum Simulations GmbH, Rintheimer Str. 23, 76131 Karlsruhe, Germany}

\date{March 2026}

\begin{document}

\maketitle

\begin{abstract}
    Simulating vibronic spectra is a central task in physical chemistry, offering insight into important properties of molecules. Recently, it has been experimentally demonstrated that photonic platforms based on Gaussian boson sampling (GBS) are capable of performing these simulations. 
    However, whether an actual GBS approach is required depends on the molecule under investigation.
    To develop a better understanding on the requirements for simulating vibronic spectra, we explore connections between theoretical approximations in physical chemistry and their photonic counterparts. 
    Mapping these approximations into photonics, we show that for certain molecules the GBS approach is unnecessary.
    We place special emphasis on the linear coupling approximation, which in photonics corresponds to sampling from multiple coherent states. 
    By implementing this approach in experiments, we demonstrate improved similarities over previously reported GBS results for formic acid and identify the particular attributes that a molecule must exhibit for this, and other approximations, to be valid.
    These results highlight the importance in forming deeper connections between traditional methods and photonic approaches.
\end{abstract}

\section*{Introduction}
Simulating vibronic spectra is one of the cornerstone tasks in physical chemistry, playing a critical role in understanding fundamental properties of molecules \cite{condon1926theory, BornOppenheimer1927, herzberg1933schwingungsstruktur}. 
For instance, it allows one to predict their luminescent properties, which is important for guiding the synthesis of more efficient organic light-emitting diodes \cite{Tuckova2022INVEST}. 
Due to its importance in both science and industry, more efficient methods for solving this task are highly sought after.

In 2015, Huh \textit{et al.} mapped the vibronic spectra simulation task into a photonic platform; the Gaussian Boson Sampling (GBS) architecture \cite{huh2015boson,Huh2017VibronicGBS}.
They showed that vibronic transitions can be mapped to a combination of coherent states, interferometers and squeezed states of light (or simply squeezing) which are the building blocks of GBS. 
Recently, Zhu \textit{et al.} have shown the first experimental implementation of vibronic spectra simulations in GBS \cite{zhu2024large}.
These results showed that the system was indeed \textit{sufficient} for simulating the chosen molecules.
Here we address the question of whether the full GBS architecture is \textit{necessary}, i.e. is the full power of GBS required, for producing accurate results. 

To answer this question, we take a deeper look into the analogies between physical chemistry and photonics.
This allows us to demonstrate that for certain molecules, as long as well-known approximations used in physical chemistry hold, simplified photonic implementations can be used to produce accurate results without the need for a full GBS system. 
This treatment reveals a hierarchy of problems that can be implemented with increasing complexity in their corresponding photonic architectures. 
In particular, we show that the spectra of a certain class of molecules can be simulated accurately by simply sampling from multiple laser sources (coherent states) with various mean photon numbers.
This corresponds to the well-known linear coupling approximation.
Remarkably, this class of molecules includes formic acid, a commonly used example in the community \cite{huh2015boson,zhu2024large,Oh2024}. 
Using this molecule, we experimentally demonstrate that the simplified photonic implementation made under this assumption is able to more faithfully reproduce the vibronic spectrum than has been shown in previous demonstrations utilizing a full GBS setup \cite{zhu2024large}. This improvement is due to the significantly reduced complexity of the experimental setup and showcases the power of increasing our understanding of the connection between physical chemistry and photonic systems. 
Extending upon these studies, we also explore another approximation, the parallel approximation, which requires the addition of squeezing and yields accurate results for a second class of molecules.
Finally, we discuss other cases of molecules that necessitate going beyond these approximations, requiring the inclusion of interferometers - i.e. requiring the complete GBS architecture.

\section*{Results}

\boldsize{From chemistry to photonics}

Vibronic transitions arise from the simultaneous change in the electronic and vibrational states of a molecule, typically occurring during the absorption or emission of light. 
The vibrational modes of molecules are quantized as phonons—the bosonic quasi-particles representing quantized lattice vibrations. 
Since photons and phonons are both bosons, they can be treated using the same mathematical formalism.
This allows photons to directly simulate phonons and vice versa.

In the Franck-Condon formalism the intensity of vibronic transitions is determined by the transition probability from the initial state $|\boldsymbol{n}\rangle = | n_{1}, \dots, n_{M} \rangle $ to the final state $|\boldsymbol{m^\prime}\rangle = | m_{1}^\prime, \dots, m_{M}^\prime\rangle $, known as the Franck-Condon factor (FCF)  $FCF_{n\rightarrow m^\prime} = |\langle \boldsymbol{m}^\prime | \boldsymbol{n}\rangle|^2$ \cite{franck1926elementary, condon1926theory}.
Here \textit{M}=3\textit{N}-6 (\textit{M}=3\textit{N}-5 for linear molecules) is the number of normal modes describing molecules with $N$ atoms and $n_i$ and $m_i^\prime$ are the vibrational quantum numbers, i.e. number of phonons in the $i$-th normal mode.

\begin{figure}
    \centering
    \includegraphics[width=1.0\linewidth]{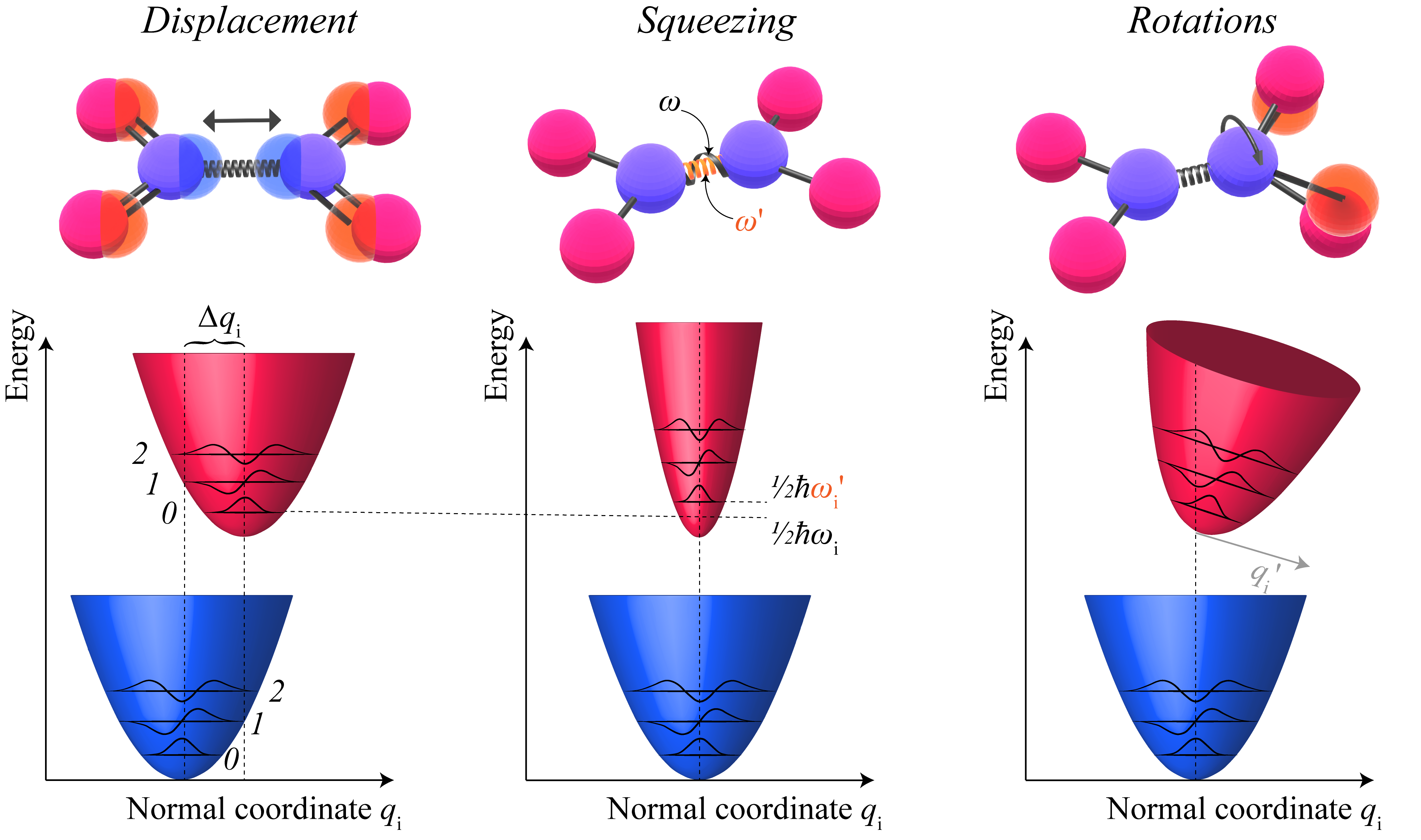}
    \caption{Pictorial representation of effects on energy levels occurring upon molecular excitation. First, a change of the geometrical structure of the molecules translates to a shift of the potential energy surface by $\Delta$\textit{$q_i$} along the normal coordinate \textit{$q_i$} (left). Second, squeezing accounts for changes in the potential energy surface leading to alteration of energy levels from $\hbar\textit{$\omega_i$}/2$ to $\hbar\textit{$\omega_i^{\prime}$}/2$ (middle). Finally, the shape of vibrations can change resulting in mixing of normal modes described by the Duschinsky rotation \textit{$q_i^\prime$}=\textbf{U}\textit{$q_i$} (right).}
    \label{Concept}
\end{figure}

For each normal mode one can define \textit{mass-weighted} normal coordinates \cite{duschinsky1937interpretation} $\boldsymbol{q} = (q_{1}, \dots, q_{M}) $ and $\boldsymbol{q^\prime} = (q_{1}^\prime, \dots, q_{M}^\prime) $ for the initial and final states, respectively. 
To transform between the normal coordinates Duschinsky proposed the relation \cite{duschinsky1937interpretation}:
\begin{equation}
\boldsymbol{q}^\prime = \mathbf{U}\boldsymbol{q} + \Delta \boldsymbol{q}.
\end{equation}
Here the displacement vector $\Delta\boldsymbol{q} =  (\Delta q_{1}, \dots, \Delta q_{M})$ is an origin shift along the normal coordinates, often corresponding to a change in the bond length of the molecule. 
It leads to the displacement of the potential energy surface (PES) along normal coordinates as illustrated in Fig.~\ref{Concept} (left).
The Duschinsky matrix $\mathbf{U}$ describes mixing between the normal coordinates, reflecting geometrical changes in the molecule and rotation of the PES as depicted in Fig.~\ref{Concept} (right).
Additionally, in the general case the vibrational frequencies \textit{$\omega_i$} and \textit{$\omega_i^{\prime}$} of the initial and final states may be different, resulting in the change of PES shape as shown in Fig.~\ref{Concept} (middle).

From the Duschinsky relation it can be shown that FCFs can be expressed using the Doktorov operator $\hat{U}_{Dok}$ as \cite{doktorov1977dynamical}
\begin{equation}
    FCF_{n\rightarrow m^\prime} = |\langle \boldsymbol{m}^\prime | \boldsymbol{n}\rangle|^2 = |\langle \boldsymbol{m} | \hat{U}_{Dok} | \boldsymbol{n}  \rangle |^2.
\end{equation}
This Doktorov operator is crucial since it transforms the final state to the same basis as the initial state $|\boldsymbol{m}\rangle = \hat{U}_{Dok} |\boldsymbol{m}^\prime\rangle$.
Its transformation can generally be described by a displacement parameter that encodes the origin shift by $\beta_i =\sqrt{\frac{\omega_i^{\prime}}{2\hbar}} \Delta q_i$ and a matrix $\mathbf{J} = \Omega^{\prime}\textbf{U}\Omega^{-1}$,
where $\Omega^{\prime} = \text{diag}\Bigl(\sqrt{\omega_1^{\prime}}, ..., \sqrt{\omega_M^{\prime}}\Bigr)$ and $\Omega = \text{diag}\Bigl(\sqrt{\omega_1}, ..., \sqrt{\omega_M}\Bigr)$ are diagonal matrices. 
The matrix $\mathbf{J}$ encodes both the frequency change ($\omega_i \rightarrow\omega^{\prime}_i$) and the mode mixing effects ($\mathbf{U}$).

To adapt the Doktorov transformation to the standard GBS scheme, Huh \textit{et al.} \cite{Huh2017VibronicGBS} used a singular value decomposition
\begin{equation} \label{eq:SVD}
\mathbf{J} = \Omega^{\prime}\textbf{U}\Omega^{-1} = \mathbf{U_L}\mathbf{\Sigma}\mathbf{U_R},
\end{equation}
rewriting $\mathbf{J}$ in terms of two unitary matrices $\mathbf{U_L}$ and $\mathbf{U_R}$ and a diagonal matrix $\mathbf{\Sigma}$ of the singular values.
With this decomposition, the Doktorov operator can be written as $\hat{U}_{Dok} = \hat{D}(\boldsymbol{\beta}) \hat{R}(\mathbf{U_L})\hat{S}(\Sigma)\hat{R}(\mathbf{U_R})$, where $\hat{D}$, $\hat{S}$ and $\hat{R}$ are displacement, squeezing and rotation operators, directly corresponding to the operations of GBS \cite{huh2015boson}. A detailed expression of these operators is given in Methods.
The displacement operator displaces the $i$-th mode by $\beta_i$, representing the origin shift of the molecule.
Squeezing is applied with a squeezing parameter of $r_i = \ln{\left(\Sigma_{i,i}\right)}$ for the $i$-th mode, and is related to both the frequency change and Duschinsky mixing. 
The two rotation operators perform a rotation described by the unitaries $\mathbf{U_L}$ and $\mathbf{U_R}$, arising from the Duschinsky mode mixing. Such a rotation corresponds to the interference between different optical modes.

In this paper we will consider only the zero-temperature approximation, which implies that we only account for transitions starting from the zero vibrational quantum number \textit{$n_i$}=0 for all modes, i.e. vacuum $|\boldsymbol{0}\rangle$. 
Hence, we can drop $\hat{R}(\mathbf{U_R})$, since $\hat{R}(\mathbf{U_R}) | \boldsymbol{0}  \rangle =  | \boldsymbol{0}  \rangle$. 
This results in a Doktorov operator written as
\begin{equation}\label{eq:Doktorov2}
			\hat{U}_{Dok} = \hat{D}(\boldsymbol{\beta}) \hat{R}(\mathbf{U_L})\hat{S}(\mathbf{\Sigma}).
\end{equation}
In this form the link between the Duschinsky approximation in chemistry and the GBS system becomes clear. 
This lays the foundation for mapping different approximations from physical chemistry into photonic implementations.

\boldsize{Linear coupling approximation}

The simplest approximation to obtain FCFs is the linear vibronic coupling model \cite{frank1975electron}. 
In physical chemistry this is also known as the Independent Mode Displaced Harmonic Oscillator (IMDHO) model \cite{KarasuluApproximations}. 
The vibrational frequencies of the initial and final states are assumed to be equal ($\omega_i = \omega_i^{\prime}$) and mode mixing is neglected, i.e. the Duschinsky matrix is assumed to be identity ($\mathbf{U} = \mathbf{I}_M$). 
In this case, the singular-value decomposition in equation~\eqref{eq:SVD} will result in an identity operation $\mathbf{U_{L}} = \mathbf{I}_M$ and squeezing parameters of $r_i = \ln{(\Sigma_{i,i}) = 0}$. 
Only the displacement operator remains and the FCFs simplify to products of Poisson distributions given by
\begin{equation}\label{eq:LinearCoupling}
		FCF_{0\rightarrow m} =  |\langle \boldsymbol{m}| \hat{D}(\boldsymbol{\beta}) |\boldsymbol{0}  \rangle  |^2
 =  \prod_{i=1}^{M}\frac{|\beta_i|^{2m_i}}{m_i!}e^{-|\beta_i|^{2}}.
\end{equation}
This approximation is well-known in physical chemistry, usually given in terms of the Huang-Rhys factor $S_i = |\beta_i|^2$ associated with a normal mode $q_i$ \cite{frank1975electron}. 
The Huang-Rhys factor describes the strength of coupling between electronic and vibrational states in a molecule.

\boldsize{Parallel approximation}

The next level of complexity beyond the linear coupling model is the parallel approximation, also known in physical chemistry as the Independent Mode Displaced Harmonic Oscillator with Frequency Alteration (IMDHO-FA) model \cite{KarasuluApproximations}. 
Both the displacement and the change in vibrational frequencies between the ground and excited states are considered, while Duschinsky rotations are still neglected ($\mathbf{U} = \mathbf{I}_M$). 
The singular-value decomposition in Eq.~\eqref{eq:SVD} will again yield the identity operation $\mathbf{U_{L}} = \mathbf{I}_M$. 
However, the singular values correspond to $\Sigma^* = \Omega^{\prime}\Omega^{-1}$, resulting in squeezing parameters of $r_i = \ln \left( \sqrt{\frac{\omega^{\prime}_i}{\omega_i}} \right)$.
This allows to write the FCFs as
\begin{equation}\label{eq:Parallel}
		FCF_{0\rightarrow m} =  |\langle \boldsymbol{m}| \hat{D}(\boldsymbol{\beta})\hat{S}(\Sigma^*) |\boldsymbol{0}  \rangle  |^2.
\end{equation}
This expression describes the probability of finding the \textit{$|\boldsymbol{m}\rangle$} photon pattern in multiple independent displaced squeezed states, for which an analytical solution can be derived \cite{gerry2005introductory} (see Methods for details).

Note that one could also make an approximation that considers Duschinsky mixing but neglects frequency change. In a photonic implementation this requires a complete GBS system and is therefore not considered. More details can be found in Methods.

\boldsize{Validity of the vibronic approximations}

It is essential to note that the above theory on vibronic coupling in molecules has been formulated within a small harmonic vibrations \cite{wilson1980molecular} approximation neglecting anharmonic effects. In addition, possible interaction between electronic states as well as post FC effects are further neglected \cite{bersuker1989vibronic}. Despite that the FC picture is sufficient for many routine applications even in its simplest form of linear coupling model \cite{Baiardi2016}. It is, however, hard to derive any consistent rules allowing to predict which approximation is sufficient for a particular system: it depends not only on the molecule, but also on the electronic state of interest as well as on the kind of experiment. Therefore, we have tested a selection of diverse molecules observed in several types of spectroscopy (see Methods and Supplementary).

\boldsize{Photonic implementations}

The core of simulating vibronic spectra with photonics is to sample photon-number patterns $|\boldsymbol{m}\rangle = | m_{1}, \dots, m_{M}\rangle$ from a multi-mode state.
Therefore, an experimental requirement, independent of the targeted approximation, is a detection stage with photon-number resolution (PNR) for all modes. 
With a pulsed laser system one can record $N_{S}$ samples by collecting the photon numbers for each pulse. 
The FCFs can then be expressed as
\begin{equation}
FCF_{0\rightarrow m} = |\langle \boldsymbol{m} | \hat{U}_{Dok} | \boldsymbol{0}  \rangle |^2 \approx \frac{N(\boldsymbol{m})}{N_{S}},
\end{equation}
where $N(\boldsymbol{m})$ is how often the pattern $|\boldsymbol{m}\rangle$ occured in the $N_S$ samples.
The spectrum, which we call the Franck-Condon profile (FCP), can then be reconstructed using \cite{huh2015boson}
\begin{equation}
FCP(\omega^{\prime\prime}) = \frac{1}{N_{S}} \sum_\mathbf{m} N(\boldsymbol{m}) \delta\left(\omega^{\prime\prime}-\sum_{i=1}^M m_i \omega_i^{\prime}\right),
\end{equation}
where $\omega^{\prime\prime}$ is the frequency of the transition, which can be calculated from the photon numbers $m_i$ and final frequencies $\omega_i^{\prime}$ as expressed in the Dirac delta function.

\begin{figure}[ht]
    \centering
    \includegraphics[width=1.0\linewidth]{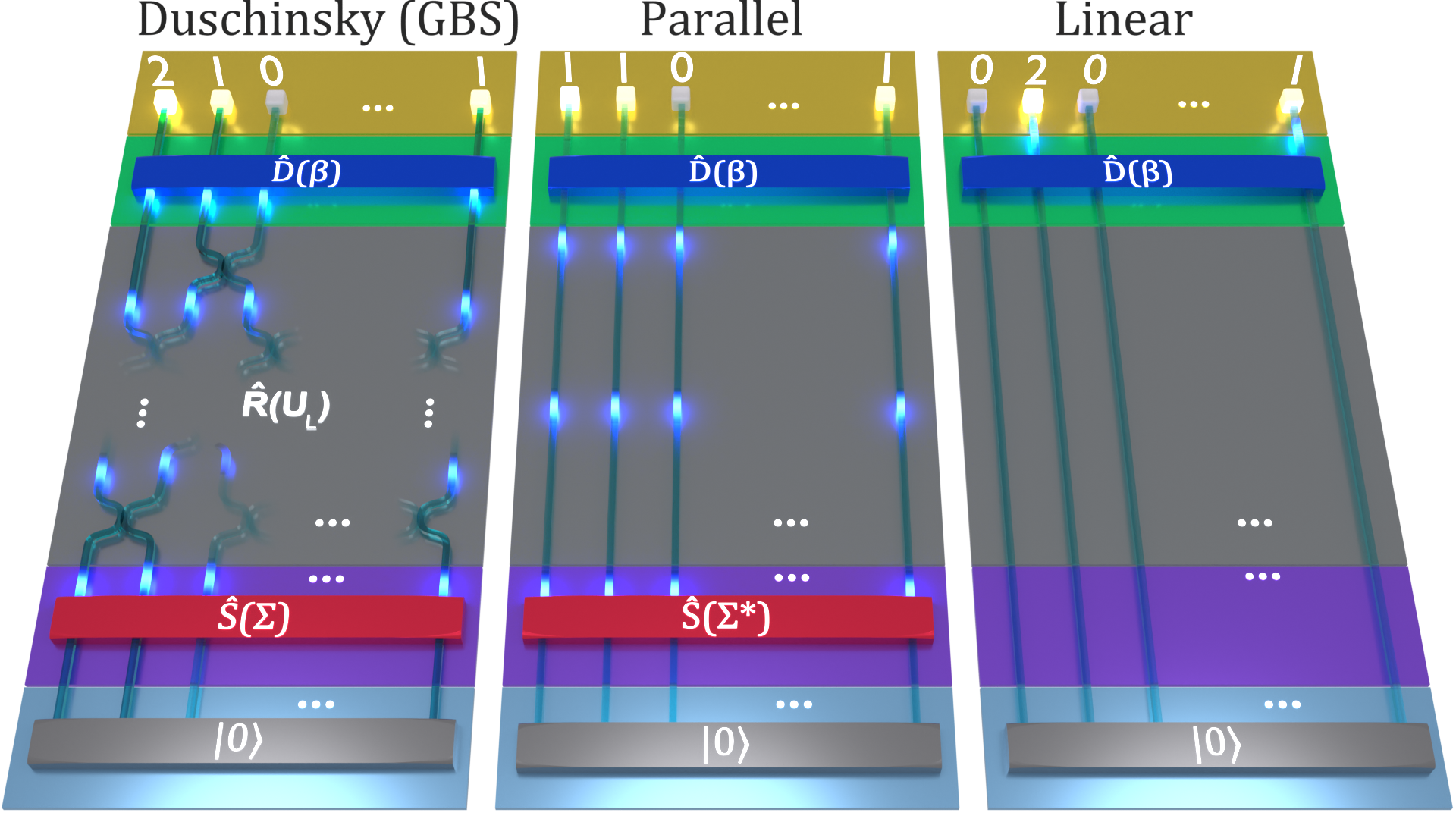}
    \caption{General schematic for implementing Duschinsky (left), parallel (middle) and linear coupling (right) approximations using photonics. Note that the displacement is applied after the interferometer, while Huh \textit{et al.} shifted it to the beginning \cite{huh2015boson} and Zhu \textit{et al.} encoded the displacement by using additional squeezed vacuum modes \cite{zhu2024large}.}
    \label{fig:Photonics}
\end{figure}

The photonic implementation of each approximation differs in the multi-mode state preparation, embodied by $\hat{U}_{Dok}$. 
For the Duschinsky approximation shown in Fig.~\ref{fig:Photonics} (left), single-mode squeezed vacuum states are sent into a multi-mode interferometer, before applying the displacement. 
This implementation requires synchronized phases at the inputs of the interferometer, as well as phase locking between the outputs and the displacement operation (see Methods).
In the simpler parallel approximation shown in Fig.~\ref{fig:Photonics} (middle), there is no need for an interferometer. 
However, phase-stability between the squeezed states and the displacements still has to be ensured.
In the linear coupling approximation only displacement is required as can be seen in Fig.~\ref{fig:Photonics} (right), making it easily implementable if one has access to PNR detectors. In addition, the system is no longer sensitive to any phase drifts and optical loss can be compensated by adjusting the displacement values $\beta_i$ accordingly. A detailed discussion on optical loss and related compensation is reported in Methods. 

Therefore, implementing parallel approximation experimentally comes with the challenges of phase noise and (non-compensable) loss on the squeezing. When implementing Duschinsky/GBS, the interferometer can also be a noise source due to imperfect unitary implementations, and especially in the case of reprogrammable systems adds a significant amount of loss as well.
For molecules that can be well described by simpler approximations, these additional experimental challenges could potentially lead to worse results compared to directly implementing the approximations.
Additionally, if one wants to benchmark their GBS system with vibronic spectra simulations, one should choose a molecule where the effects of squeezing and interferometer are strongly pronounced as the linear coupling and parallel approximations serve as a benchmark for validating the system performance.

\begin{figure}
    \centering
    \includegraphics[width=1.0\linewidth]{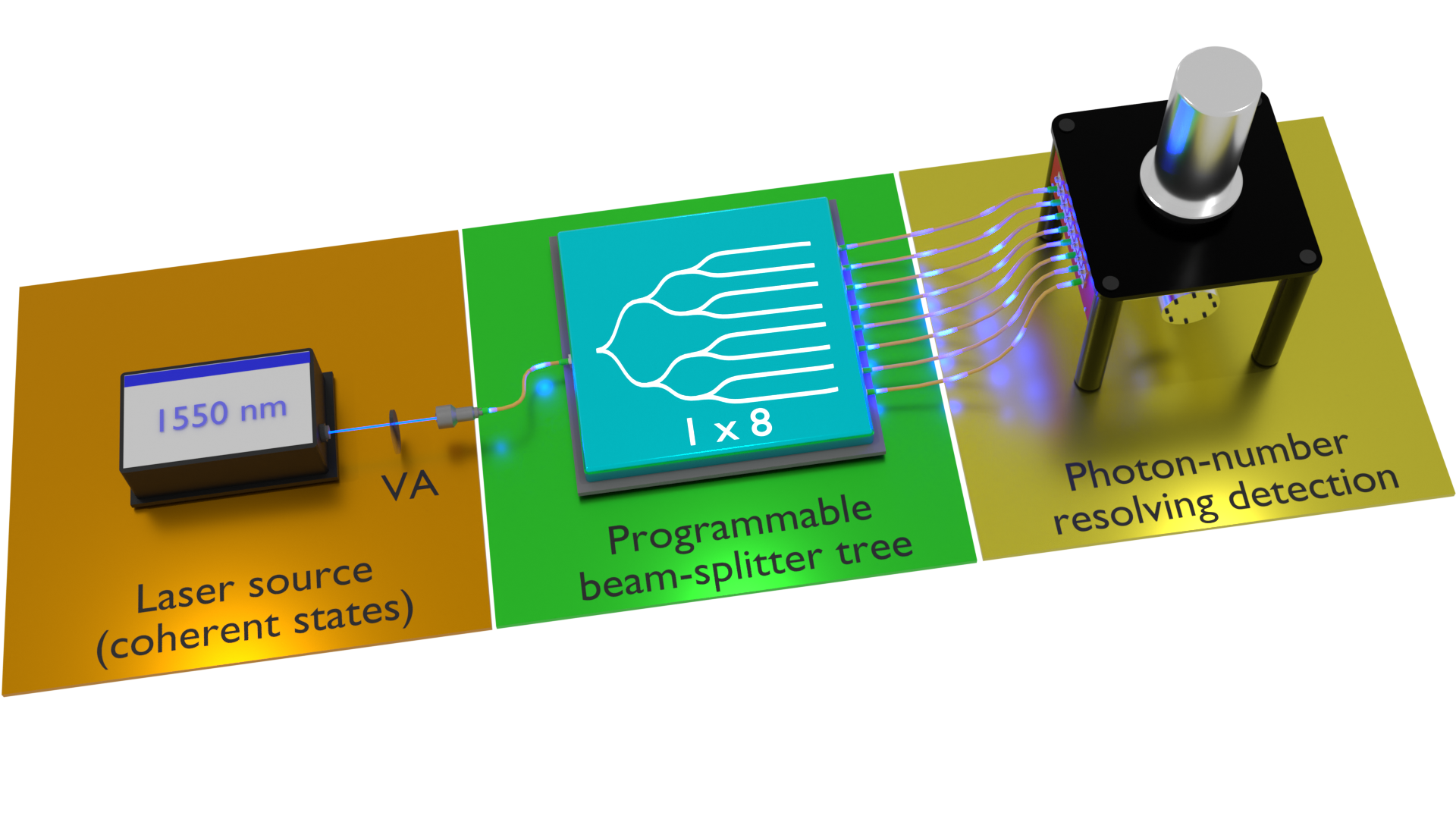}
    \caption{Schematic of the experimental setup for sampling vibronic spectra under linear coupling approximation. A source of coherent states (laser) is attenuated by a variable attenuator (VA) to reduce the overall mean photon number. The programmable beam-splitter tree splits the beam into multiple outputs with specific splitting ratios to set the required mean photon number for all the vibronic modes (conceptually shown). Photon-number resolving detectors are used to measure the exact number of photons for every sample (pulse).}
    \label{fig:SetupFigure}
\end{figure}

Fig.~\ref{fig:SetupFigure} shows a simplified schematic of our setup for realizing measurements made under the linear coupling approximation.
A pulsed laser at $1550$~nm (Menlo Systems GmbH, SmartComb) is coupled to a programmable photonic processor (QuiX Quantum) implementing a variable 1x8 beam-splitter tree. This allows to create a source of 8 coherent states with different mean photon numbers.
Superconducting nano-wire single-photon detectors (Single Quantum) are used to measure the photon number in each pulse at the outputs. With those we can achieve an intrinsic PNR of up to $3$ photons with high confidence \cite{IntrinsicPNR}. 
Furthermore, for modes with $|\beta_i|^2\geq0.35$ we used a time-space multiplexed detection \cite{Achilles2003} with $16$ bins, allowing to resolve up to $40$ photons in one vibronic mode (not shown in Fig.~\ref{fig:SetupFigure}, see Supplementary for details on the setup). 
In the case of molecules with more than $8$ modes, or multiple modes with $|\beta_i|^2\geq0.35$, the measurements were split up in multiple runs and combined afterwards since the individual modes are independent (see Methods for details).

\boldsize{Benchmarking the approximations}

We measured the vibronic spectra of multiple molecules in the linear coupling approximation using our photonic setup. 
For each experimental run $18\times10^7$ samples were taken.
Additionally, we performed numerical simulations for the parallel approximation and GBS approach, generating $18\times 10^7$ and $2.5\times10^6$ samples, respectively. 
We highlight that the GBS approach corresponds to the Duschinsky approximation and therefore serves as a benchmark. Note that the reduced number of samples in the latter case is due to the computational complexity of GBS \cite{Hamilton2017, KruseGBS} demanding a significant amount of computational resources and computation time. 
For simulating GBS we use The Walrus library \cite{Gupt2019}.
The validity of the simulation was verified by comparing the results to classically simulated spectra using FCClasses \cite{FCClasses}.
More details on the measurements, selected molecule parameters (description of spectroscopy types and origin of parameters) and simulations can be found in the Supplementary.

\begin{figure}
    \centering
    \includegraphics[width=\textwidth]{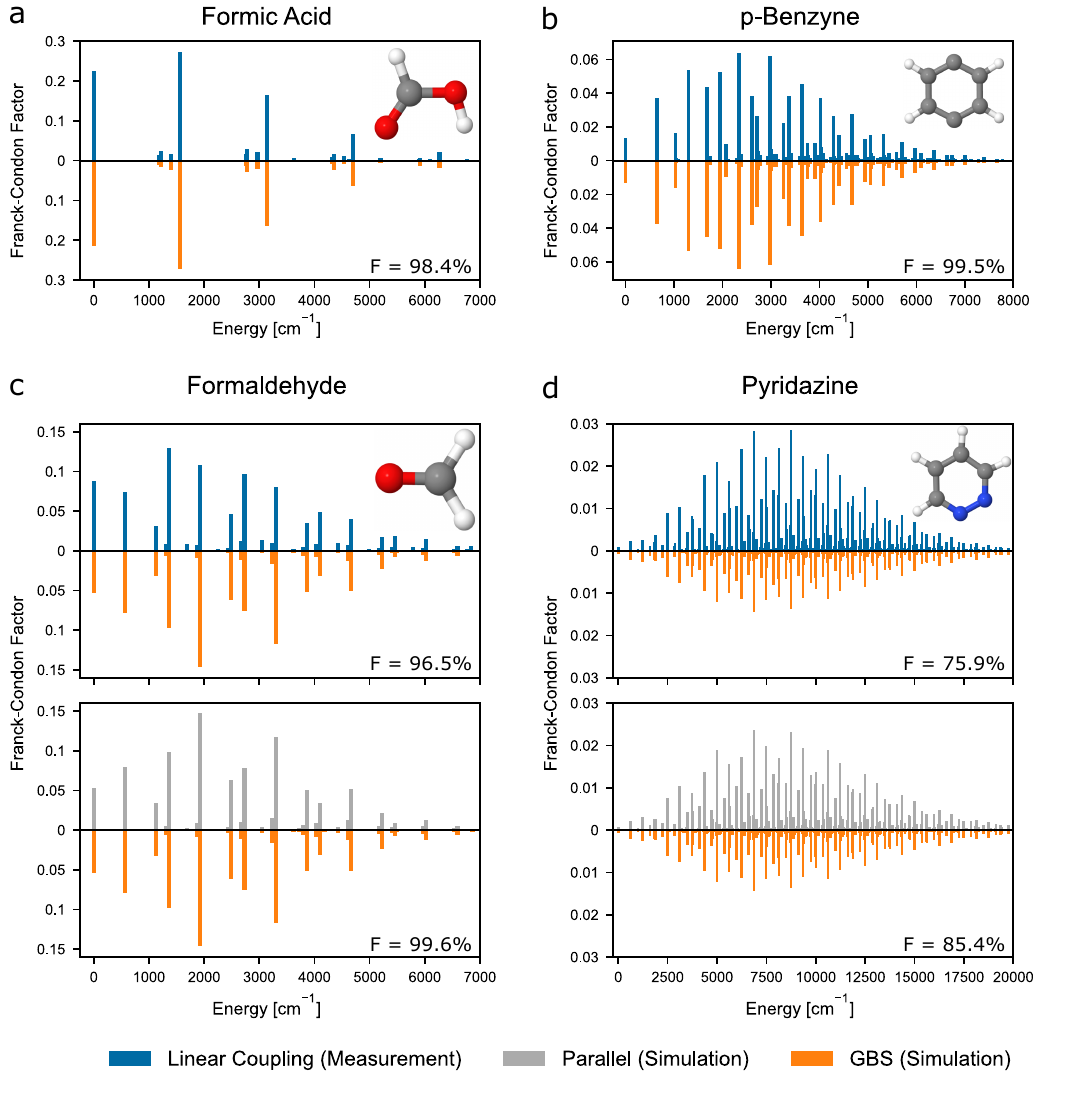}
    \caption{Measurement results for vibronic spectra in linear coupling approximation (blue) and simulation results for parallel approximation (gray) compared to the Duschinsky/GBS simulation (orange) as a benchmark. The insets show the geometric structures of the molecules in their initial electronic states: gray - carbon; white - hydrogen; red - oxygen; blue - nitrogen.
    Formic acid (a) and p-benzyne (b) show a similarity of $98.4$~\% and $99.5$~\% between linear coupling and GBS, respectively.
    In the case of Formaldehyde (c), although reaching a similarity of $96.5$~\%, linear coupling fails to show the same structure for the spectrum. However, the parallel approximation is sufficient for this molecule and reaches a similarity of $99.6$~\%.
    For pyridazine (d) neither linear coupling nor parallel approximation correctly identify the values of the FCFs, yielding similarities of $75.9$~\% and $85.4$~\% respectively, therefore requiring the GBS approach.
    Note that the plots only show $FCFs \geq 0.005\cdot FCF_{max}$ for the sake of visibility.
    Assuming Poissonian errors, the uncertainty on the similarities is $0.04$~\% for all molecules.
    A short summary for the choice of molecular parameters can be found in the Methods section. More details on molecular parameters as well as a detailed discussion of the impact of imperfect displacements on the measured similarities are given in the Supplementary.}
    \label{fig:results}
\end{figure}

To compare the results for the different photonic implementations, we use the similarity
$F = \sum_{\omega^{\prime\prime} = 0}^{\infty}\sqrt{p(\omega^{\prime\prime})\cdot q(\omega^{\prime\prime})}$,
where $p(\omega^{\prime\prime})$ and $q(\omega^{\prime\prime})$ are the FCFs of the two spectra to be compared.
The measurement and simulation results for a subset of molecules are shown in Fig.~\ref{fig:results}.
We can identify that our photonic measurement in the linear coupling approximation accurately reconstructs the vibronic spectra of formic acid (Fig.~\ref{fig:results} a) and p-benzyne (Fig.~\ref{fig:results} b) with $98.4$~\% and $99.5$~\% similarity, respectively. 

Of particular importance is the case of formic acid, which is a commonly chosen example for simulating molecular vibronic spectra using  GBS systems \cite{huh2015boson, zhu2024large, Oh2024}. The similarity of $92.9$~\% Zhu \textit{et al.} achieved using a full GBS system \cite{zhu2024large} is well below the $98.4$~\% achieved here, further noting that this is despite the fact that we have considered a greater number of modes (7 instead of 4).
Our result highlights that for molecules where the linear coupling approximation is valid, the simplified scheme can even outperform GBS.
This is especially the case for molecules where Duschinsky rotation effects are almost negligible, as shown here for formic acid and p-benzyne.

We believe that formic acid was chosen as a common example due to considerable mode mixing that can be seen in the Duschinsky matrix (see Supplementary), which seems to contradict our result. 
However, since the displacement is applied after squeezing and interferometer, as shown in Fig.~\ref{fig:Photonics}, the Duschinsky mixing will only alter the spectrum significantly if the squeezing parameters of the corresponding modes are non-negligible.
In the case of formic acid these are negligible, so that Duschinsky mixing affects only weak progressions around 1400 $cm^{-1}$ and 3000 $cm^{-1}$. 

For formaldehyde (Fig.~\ref{fig:results} c) it can be observed that, although reaching a rather high similarity of $96.5$~\%,  the linear coupling approximation leads to an altered spectrum when compared to Duschinsky. 
Adding squeezing (considering frequency change) via the parallel approximation corrects this shape mismatch, resulting in a similarity of $99.6$~\%. The required squeezing parameters in that example are $\mathbf{r} \approx (-0.379,\,-0.173,\,-0.081,\,-0.143,\,0.021,\,0.028)$, corresponding to a maximum squeezing of about $3.3$\,dB, which can be easily reached in experiments.
Therefore, formaldehyde could be accurately simulated in a photonic platform with the use of displaced squeezed states and would not require an interferometer to implement rotations. This approach may be able to outperform GBS systems as well, due to the reduced experimental effort.

The last example is pyridazine (Fig.~\ref{fig:results} d), where neither linear coupling ($75.9$~\%) nor parallel approximation ($85.4$~\%) lead to accurate reconstruction of the vibronic profile. 
While some changes in shape of the spectrum are present, they are difficult to spot by eye.
They are mostly present in changes of peak ratios, for example the ratio between the two maximum peaks $\frac{FCF(\omega''\approx 8740.16)}{FCF(\omega''\approx 6862.51)}$ drops from $1.01$ for linear coupling and $0.98$ for parallel approximation to $0.95$ for the GBS simulation, meaning the maximum becomes slightly more pronounced.
The significantly more pronounced change between the approximations and the GBS simulation is notable in terms of the actual Franck-Condon factors rather than the shape of the spectrum.
This change in the overall FCFs is mostly caused by the Duschinsky mixing redistributing the probabilities from more significant peaks to previously weakly populated or completely unpopulated energy levels, leading to a broad background contribution in the spectrum.
For applications where precise FCF values are required, for example in the calculation of branching ratios \cite{BranchingRatios}, one therefore needs the complete GBS system to simulate the vibronic spectrum of pyridazine in a photonic platform.
In that case the maximum squeezing one would have to produce is $|r|_{max}\approx0.651$, which is around $5.7$\,dB squeezing and still experimentally reachable.

It is instructive to compare vibronic structures of the band in pyridazine \cite{fischer2000vibronic} with chemically similar molecules: naphthalene \cite{ferguson1957vapor} and para-benzyne \cite{wenthold1999negative}. Although all compounds are rigid aromatic cycles, only pyridazine exhibits the need for higher-order approximations. This can be \textit{a posteriori} explained by the lower symmetry of the pyridazine molecule which allows for extensive vibrational mode mixing due to less symmetry constraints.

We have considered a wider molecular set (see Supplementary). However, for most molecules data analysis and simulations show that spectra are well described by the linear coupling model. 
This confirms that Duschinsky mixing and frequency change are not ubiquitous \cite{frank1975electron,Baiardi2016} and highlights the importance of carefully choosing molecules for full GBS implementations.

\section*{Discussion}
We have shown that in physical chemistry well-known approximations for molecular vibronic spectra simulations can be directly related to photonic implementations, allowing one to substantially decrease the complexity of the photonic system required to simulate certain classes of molecules. The commonly used linear coupling approximation, which only assumes displacement of nuclei, can be performed on photonic systems by sampling the photon number from coherent states. 
Furthermore, including frequency change of the vibronic modes, known as the parallel approximation, is analogous to sampling photon numbers from displaced squeezed states. 
In order to obtain a high similarity between experimental results and the physical spectra, the GBS system is only required when these two cases are not valid.  

Through understanding these different degrees of complexity, we have shown that the vibronic spectra of many molecules, including formic acid, can be accurately reconstructed with the linear coupling model. 
This example is of particular interest, as it is commonly used to demonstrate the capabilities of the GBS approach \cite{huh2015boson, zhu2024large, Oh2024}.
Additionally, experimental imperfections present in GBS systems such as optical loss on squeezed states \cite{oh2024classical} or imperfections in the implemented interferometer will inevitably degrade the measurement outcome and limit the accuracy of the obtained vibronic spectra \cite{clements2018approximating}.
With our simplified photonic setup one can easily compensate for loss or imperfect splitting ratios by adjusting the mean photon numbers at the detection stage. The decreased experimental complexity leads to improved similarities when comparing to the previous implementation \cite{zhu2024large}.

Furthermore, we have presented an example (formaldehyde) from a second class of molecules, for which the slightly more complex parallel approximation produces accurate spectra. 
This highlights that only checking if the linear coupling approximation fails is not sufficient to select molecules for GBS implementations.

Finally, with pyridazine we also provide an example of a molecule where neither linear coupling nor parallel approximations are sufficient to obtain accurate results. Pyridazine is therefore an example of a molecule for which one requires the complete GBS system in order to accurately determine its vibronic spectrum.
The presented hierarchy of approximations, and corresponding examples, provide an important guideline for choosing molecules in GBS implementations.

As a final note, we want to address the topic of quantum advantage for vibronic spectra simulations in GBS. 
Most recently, this question was investigated in detail by Oh \textit{et al.} \cite{Oh2024} and Lim \textit{et al.} \cite{Lim2025}. 
Oh \textit{et al.} concluded that vibronic spectra simulations with the standard GBS scheme will not reach quantum advantage, but that it may still be possible by utilizing squeezed Fock states as an input resource \cite{Oh2024}. 
Lim \textit{et al.} recently investigated this proposal and concluded that this scheme will also not yield quantum advantage \cite{Lim2025}.
Nevertheless, deeper understanding of such interdisciplinary connections, as presented here, ensures that progress and insight achieved in one field is translated into the second, and provides a stronger foundation for future investigations.

\newpage
\section*{Methods}

\boldsize{Optical operators}

The Doktorov operator is decomposed into displacement $\hat{D}(\boldsymbol{\beta})$, rotation $\hat{R}(\mathbf{U_{L}})$ and squeezing $\hat{S}(\mathbf{\Sigma})$ operations to map it into a photonic approach, as shown in equation~\eqref{eq:Doktorov2}.
The detailed expressions of these multi-mode operators are given by
\begin{equation}
    \hat{D}(\boldsymbol{\beta}) = \prod_{i=1}^M \exp\biggr[\beta_i \hat{a}_i^{\dagger} - \beta^{*}_i \hat{a}_i\biggl],
\end{equation}

\begin{equation}
    \hat{S}(\mathbf{\Sigma}) = \prod_{i=1}^M \exp\biggr[\frac{r_i}{2} (\hat{a}^{\dagger}_i)^2 - \frac{r_i^*}{2} (\hat{a}_i)^2\biggl],
\end{equation}

\begin{equation}
    \hat{R}(\mathbf{U_{L}}) = \exp\biggr[\sum_{i,j} [ln(U_{L})]_{i.j} \hat{a}^{\dagger}_i \hat{a}_j\biggl],
\end{equation}

with $\hat{a_i}$ ($\hat{a}_i^{\dagger}$) being the photon number annihilation (creation) operators for mode $i$.
From the structure of these operators it becomes clear that the displacement and squeezing operations act on individual modes, while the rotation operation is a mode-mixing process. \cite{MaOperators}

\boldsize{Analytical solution for parallel approximation}

As discussed in the main text, Franck-Condon factors in the parallel approximation (assuming zero temperature) can be calculated as
\begin{equation}\label{Parallel2}
		FCF_{0\rightarrow m} =  |\langle \boldsymbol{m}| \hat{D}(\boldsymbol{\beta})\hat{S}(\Sigma^*) |\boldsymbol{0}  \rangle  |^2,
\end{equation}
where $\hat{D}(\boldsymbol{\beta})$ is the displacement operator with $\mathbf{\beta} = (\beta_1, \dots, \beta_M)$ and $\beta_i = |\beta_i| e^{i\phi_{D,i}}$ can be expressed in terms of displacement amplitude $|\beta_i|$ and phase $\phi_{D,i}$. $\hat{S}(\Sigma^*)$ is the squeezing operator with the diagonal matrix $\Sigma^*$ and squeezing parameters $r_i = \ln(\Sigma^*_{i,i})$. The squeezing parameter can also be written in terms of squeezing amplitude $|r_i|$ and phase $\theta_i$ as $r_i = |r_i|e^{i\theta_i}$. $|\boldsymbol{m}\rangle = | m_{1}, \dots, m_{M} \rangle $ is the vibrational quantum number of the final state. The analytical solution for considering a single mode $|\langle m_i|\hat{D}(\beta_i)\hat{S}(r_i)|0\rangle|^2$ can be found in \cite{gerry2005introductory}. 
The Franck-Condon factor that we want to calculate in our approximation is then the product of photon number probabilities of multiple displaced squeezed states:
\begin{equation}\label{ParallelFULL}
\begin{aligned}
\text{FCF}_{0\rightarrow m} &= |\langle \boldsymbol{m}| \hat{D}(\boldsymbol{\beta})\hat{S}(\Sigma^*) |\boldsymbol{0}  \rangle  |^2 \\
&= \prod_{i=0}^M\frac{1}{m_i! \cosh(|r_i|)} \left( \frac{1}{2} \tanh(|r_i|) \right)^{m_i} \left| H_{m_i}\left( \frac{\gamma_i}{\sqrt{e^{i\theta_i} \sinh(2|r_i|)}} \right) \right|^2 \\
&\quad \times \exp\left( -|\beta_i|^2 - \frac{1}{2} \left( \beta_i^{*2} e^{i\theta_i} + \beta_i^2 e^{-i\theta_i} \right) \tanh(|r_i|) \right).
\end{aligned}
\end{equation}
Here $\gamma_i = \beta_i \cosh(|r_i|) + \beta_i^* e^{i\theta_i} \sinh(|r_i|)$ and $H_{m_i}$ is Hermite polynomial. 
It is possible to show that if one sets $r_i=0$ the Hermit polynomial part and its pre-factor reduce to $|\beta_i|^{2m}$ and the exponent reduces to $e^{-|\beta_i|^2}$. Hence, we arrive at the original linear coupling approximation.

From equation~\eqref{ParallelFULL} it is clear that the final photon probabilities depend on the displacement and squeezing phases $\phi_{D,i}$ and $\theta_i$, respectively. This imposes additional requirements for experimental implementation of the parallel approximation. The displacement and squeezing phases should be locked to one another at a specific angle depending on the vibrational mode. This can be challenging to implement for large number of vibrational modes.

\boldsize{Duschinsky rotation without frequency change}

For completeness, we briefly mention another approximation that can be considered - Duschinsky rotation ($\mathbf{U} \neq \mathbf{I}_M$) without frequency change ($\omega_i = \omega_i^{\prime}$).
In this approximation the singular value decomposition~\eqref{eq:SVD} will in general yield squeezing parameters of $r_i \neq 0$ and a unitary $\mathbf{U_L} \neq \mathbf{I}_M$.
Squeezing arises from the Duschinsky rotation mixing the frequencies of the different modes as seen in equation~\eqref{eq:SVD}. This frequency mixing can be interpreted as a change in frequency, leading to the requirement of a squeezing operator to represent this change.
Therefore, the scheme for the photonic implementation does not differ from the Duschinsky approximation in Eq.~\eqref{eq:Doktorov2} since the same set of operators would be needed.
Note that even if a special case with $r_i=0$ and $\mathbf{U_L} \neq \mathbf{I}_M$ could be found, in the zero-temperature approximation the rotation would be applied to the vacuum state and Eq.~\eqref{eq:Doktorov2} collapses to the linear coupling approximation.
For these reasons, this approximation is not considered in this paper.

\boldsize{Loss compensation}

One crucial advantage of coherent light is that its photon-number statistics remain Poissonian when experiencing photon loss.
In contrast, the photon statistics of squeezed light changes upon loss. 
This can be seen when looking at the photon-number statistics of a squeezed vacuum \cite{gerry2005introductory}
\begin{equation}
\hat{S}(r)|0\rangle = \sum_{m = 0}^\infty c_{2m}|2m\rangle
\end{equation}
which only consists of even photon-number components. 
When experiencing loss uneven photon number components will arise, altering the photon-number statistics.
Since the vibronic spectra simulations in photonics rely on sampling from the photon-number statistics, loss will introduce errors and inevitably reduce the quality of the result if squeezing is involved.

In the case of coherent light, loss can be compensated by adjusting the displacement parameters accordingly such that the detected displacement corresponds to what one would like to sample from. Specifically, if mode $i$ experiences a loss of $L_i$, the input displacement can be adjusted as $\beta_i^\prime = \beta_i/\sqrt{1-L_i}$.

\boldsize{Requirement of phase stability for GBS}

As mentioned in the main text, the full GBS approach requires phase locking at the inputs of the interferometer, as well as between the outputs and displacements.
It was shown that performing GBS corresponds to sampling from a specific matrix that can be constructed from the covariance matrix of the whole system \cite{Hamilton2017,KruseGBS}. 
The covariance matrix depends on the relative phases between the squeezed state inputs, showing that interference between multiple squeezed states is phase-dependent. 
Therefore, one requires phase locking between the input modes for performing vibronic spectra simulations in the system, additionally to the phase locking between squeezing and displacement explained before.

\boldsize{Molecular parameters}

Experimentally the vibronic structure of molecules can be observed using three main techniques - namely photoelectron spectroscopy, negative ion photoelectron spectroscopy (NIPES) and ultraviolet-visible (UV-vis) spectroscopy. To thoroughly benchmark approximations presented in our article, we chose molecules for each of the three spectroscopic techniques. Below one can find a quick summary of how the molecular data was obtained for the molecules that are presented in the main text.

For photoelectron spectroscopy we used formic acid as an example molecule. The data for its photonic computing implementation has been derived from \cite{jankowiak2007vibronic}, while the structure and harmonic vibrations of formic acid and its cation were optimized using CCSD(T) \cite{raghavachari1989fifth} with the cc-pVTZ basis set \cite{dunning1989gaussian} using the ORCA software package \cite{ORCA5}.

For NIPES technique we used p-benzyne as an example. For this molecule the structure optimization and calculation of harmonic vibrations were done with the complete active space self-consistent field (CASSCF) method \cite{SUN2017291}, with def2-TZVP (p-benzyne) and def2-TZVPD  \cite{B515623H} basis sets (with the corresponding resolution-of-identity auxiliary basis set) using PySCF software package \cite{sun2007python}. The active space sizes were 8 electrons in 8 orbitals for p-benzyne; 6 electrons in 6 orbitals for methylene.

Finally, we chose pyridizine for the UV-vis spectroscopy technique. The calculations are done using time-dependent density functional theory with the hybrid B3LYP \cite{becke1993new} functional using def2-TZVP \cite{B515623H} basis set.

More details for each spectroscopic technique as well as additional example molecules can be found in the Supplementary.

\boldsize{Details on measurement and data analysis}

In the case of formaldehyde, p-benzyne and pyridazine, the parameters were calculated for all vibronic modes. 
Details on that will be available in the Supplementary.
For formic acid $7$ vibronic modes using the parameters from Huh \textit{et al} \cite{huh2015boson} were measured.
For the photonic measurement only modes with displacements $|\beta_i|^2 > 10^{-4}$ were considered due to the negligible effect of smaller displacement values (probability $p(m_i>0) < 0.01$~\%) and limitations of our setup. 
Additionally, due to multiple modes with $|\beta_i|^2 \geq 0.35$ (probability $p(m_i>3) \geq 0.05$~\%) the measurements for some molecules were split into multiple runs:
\begin{itemize}
    \item formic acid: 1 measurement with 7 modes
    \item formaldehyde: 2 measurements with 2 modes each
    \item p-benzyne: 2 measurements with 5 modes each
    \item pyridazine: 3 measurements with 3 modes each
\end{itemize}
To get the full spectrum, the energies and probabilities were calculated for each measurement run separately. 
Afterwards, the whole spectrum was reconstructed by summing the energies, and multiplying the corresponding probabilities in all possible combinations.

When calculating the spectra, two energies $E_i$ and $E_j$ with $|E_i-E_j| \leq 10^{-10}$ are assumed to be equal and binned to the same energy. However, this only appears for the molecule PNA presented in the Supplementary.
The same condition for equal energies is then used in the calculation of the similarities. 

\section*{Data Availability}
The molecular parameters, optical parameters as well as the simulated and measured vibronic spectra for all considered molecules will be available at \url{https://doi.org/10.5281/zenodo.18969354}.

\section*{Code Availability}
Not applicable.

\section*{Acknowledgments}
The authors would like to thank Timon Schapeler and Tim Bartley for aid in implementing photon number-resolved detection.
M.R. would like to thank Rashid Valiev for many fruitful discussions.
This work has received funding from the German Federal Ministry of Research, Technology and Space (BMFTR) within the PhoQuant project (Grant No. 13N16103).

\section*{Author Contributions}
J.E. and V.R. conceived the presented idea.
J.E., V.R. and R.S. performed simulations and calculations.
K.L., S.A., C.P., F.L., J.L., J.E. and F.S. designed, built and/or ran the experimental setup.
J.E., V.R. and M.R. performed data analysis.
J.E., M.R. and V.R. wrote the article with input from other authors.
C.S., M.S. and B.B. directed the research.
All authors edited the paper.

\section*{Competing Interests}
The authors declare no competing financial or non-financial interests.

\bibliographystyle{naturemag.bst}
%\bibliography{vibronic_refs.bib}
\bibliography{References2.bib}

\end{document}

\maketitle

\section{Franck-Condon theory}
   
    In this work, we restrict ourselves to the Franck-Condon picture of vibronic transitions \cite{franck1926elementary, condon1926theory}. It assumes that the adiabatic approximation holds (\textit{i.e.} no vibronic coupling between electronic states exists) and that the electronic transition occurs instantaneously. Under these assumptions, the total wave function is a product of electronic $\psi_A$  and vibrational wave functions: $\chi_{An}$.  The probability amplitude for a transition between states $| \psi_A \chi_{An} \rangle$ and $| \psi_B \chi_{Bm} \rangle$ is then defined as a matrix element of the dipole moment operator $\hat{\mu}$:

    \begin{equation} \label{FC}
        P = \langle \psi_B \chi_{Bm} | \hat{\mu} |  \psi_A \chi_{An} \rangle \approx \langle \psi_B | \hat{\mu} | \psi_A  \rangle \langle \chi_{Bm} | \chi_{An} \rangle.
    \end{equation}
    
    In this equation, the first term is the permanent electronic dipole moment independent of the vibrational states $m$ and $n$, while the second term is the overlap between vibrational wave functions within different electronic states $A$ and $B$. This term is known as the \textbf{Franck-Condon factor (FCF)}, often defined as the square of the overlap. FCFs define the fine structure of the electronic transition from $A$ to $B$. Equation~\eqref{FC} is not exact as we have neglected higher-order terms, such as the Herzberg-Teller contribution \cite{herzberg1933schwingungsstruktur}, which is beyond the scope of the treatment presented here.  In this work we focus on computing Franck-Condon factors within the harmonic oscillator approximation for vibrations treated with the classical Wilson's GF method \cite{wilson1980molecular}. This assumes that molecular potential energy surface is approximated by a quadratic potential. For polyatomic molecules, molecular vibrational wave functions factorize as a products of independent harmonic oscillator wave functions, each corresponding to a vibrational normal mode with a characteristic vibrational frequency. Normal coordinates diagonalize the force constant matrix, which leads to factorization. In addition, we restrict ourselves to zero temperature, i.e. initial vibrational states are always ground ones, $n=0$. This is a reasonable assumption for low-temperature experiments (including room-temperature ones), where most molecules are in their vibrational ground states.

\section{Linear vibronic coupling model}

\subsection{Diatomic molecules}
	The simplest approximation to compute vibronic spectra in the Franck-Condon approximation within the harmonic approximation is the \textbf{linear coupling model} \cite{frank1975electron}. It assumes that vibrations have the same frequency and shape (normal modes) in the initial and final electronic states. The only difference is the origin shift, corresponding to a difference in the geometric structures in the two electronic states.
	
	For simplicity, let us consider a diatomic molecule with one vibrational normal mode with reduced mass of $M$, non-mass-weighted normal coordinate $Q$ (mass-weighted coordinate of $q = \sqrt{M} Q$) and vibrational frequency $\omega$ identical in both electronic states within the linear coupling model. The harmonic wave functions in the electronic state $A$ (vibrational state $n$) and $B$ (vibrational state $m$) displaced with respect to one another are given by:
	\begin{align}
		\chi_{An}= \left(\frac{M \omega}{\pi \hbar}\right)^{1/4} \frac{1}{\sqrt{2^n n!}} H_n(\xi) e^{-\xi/2}; \xi = \sqrt \frac{M \omega}{\hbar} Q \\
		\chi_{Bm}= \left(\frac{M \omega}{\pi \hbar}\right)^{1/4} \frac{1}{\sqrt{2^m m!}} H_m(\eta) e^{-\eta/2}; \eta = \sqrt \frac{M \omega}{\hbar} (Q - \Delta Q) .
	\end{align}

	In these equations, $H_n$, $H_m$ are Hermite polynomials and $\Delta Q$ is the geometry difference (origin shift) between the two electronic states in normal coordinates. The mass-weighted displacement parameter of the Duschinsky relation is $\Delta q = \sqrt{M} \Delta Q$. 
 
 It is convenient to introduce the dimensionless displacement parameter:
	\begin{align}
		\Delta = \sqrt \frac{M \omega}{\hbar} \Delta Q = \sqrt \frac{\omega}{\hbar} \Delta q ;\\
		 \eta = \xi - \Delta.
	\end{align} 
	 
	Assuming only transitions from the ground vibrational state of electronic state $A$, $n=0$, (\textit{i.e.} zero temperature) we can work out the FCF analytically:
	
	\begin{equation}\label{Poisson}
		F_{m0} = |\langle \chi_{Bm} | \chi_{A0}  \rangle |^2 = \
		 \frac{(\Delta ^2/2)^m}{m!}e^{-\Delta^2/2} =  \frac{S^m}{m!}e^{-S},
	\end{equation}
	where $S = \Delta^2/2$ is the Huang-Rhys factor \cite{Huang1950TheoryOL}. One can show that Huang-Rhys factor is proportional to the molecular reorganization energy upon vibronic transition, expressed in the units of the vibration frequency:
	\begin{equation}
		E_R = S \hbar \omega.
	\end{equation}
	Since in the harmonic approximation vibrational levels are equidistant, $S$ is equal to the number of vibrational quanta in state $\chi_{Bm}$, or the level of vibrational excitation.
	
	We note here that equation~\eqref{Poisson} is also a Poisson distribution. These results are trivially generalized to the case of polyatomic molecules with $3N-6$ (or $3N-5$ for linear molecules) vibrational modes, where $N$ is the number of atoms. The vibronic spectrum is computed as a convolution of $3N-6$ ($3N-5$) Poisson distributions. This  latter can be done deterministically, or sampled stochastically.
    
\subsection{A note on polyatomic molecules}
In polyatomic molecules, normal coordinate systems are, in general, different for various electronic states. Therefore, one can utilize normal coordinates of either the initial or the final state. When the coordinate system differences are taken into account by the Duschinsky theory, the result is independent of the choice of the working reference frame. Conversely, within the linear coupling model (as well as within the parallel approximation) the computed FCFs do depend on the coordinate system. 

It is often the case that the linear coupling model is applied when the vibrational spectrum is available only for the initial state: in the case of electronic excitations, computation of vibrations in the ground state is technically simpler and faster than in the excited state. In such cases the linear coupling model based on initial (ground) state vibrational data is the only option. However, if the vibrational problem has been solved for both electronic states it is recommended to use the normal coordinate frame and the frequencies of the final (excited) state \cite{koziol2009ab}. The reason is the fact that the application of the Condon principle results in vibrational excitations in the final electronic state. This is particularly true for the zero-temperature case, where only the vibrationally ground state in the initial electronic state makes contributions to the FCF. \textbf{In this work we use normal coordinates of the final state in both linear coupling and parallel regimes}.

    \section{Quantum Chemistry and Vibronic Spectroscopy}
    Vibronic structure of spectral bands can be observed in different spectroscopic techniques. Most generally electronic state of a molecule changes upon photon absorption/emission and electron attachment/loss. The former category includes absorption spectroscopy in UV/visible region \cite{atkins2014atkins}, whereas the latter one encompasses photoelectronic spectra \cite{atkins2014atkins} and negative ion photoelectronic spectra (NIPES) \cite{wenthold1999negative}. To test the linear coupling model, we simulated several molecules for each of the three techniques above as summarized in Table~\ref{table:overview}. The selection is based on availability of experimental observations, although comparison between our simulations and spectroscopic measurements is beyond the scope of this work. 
    
    For photoelectron spectroscopy and NIPES, the molecules are in the ground electronic state of the specified multiplicity. For the former, the initial state is charge-neutral, whereas the final one is positively charged (cation). For NIPES the initial state is negatively charged (anion), while the final state is neutral. For UV-vis spectroscopy the molecules are charge-neutral.
    Omitting the multiplicity, the three techniques can be summarized in a simplified manner, where $A^{n}$ is the molecule of charge $n$, $^{*}$ denotes electronically excited state, $\hbar \omega$ is photon, and $e^{-}$ is free electron:
    \begin{itemize}
        \item photoelectron: $A^{0} + \hbar \omega \rightarrow A^{+} + e^{-}$;
        \item NIPES: $A^{-1} + \hbar \omega \rightarrow A^{0} + e^{-}$;
        \item UV/vis absorbtion: $A^{0} + \hbar \omega \rightarrow A^{0*}$.
    \end{itemize}
    In all three techniques incoming photons are absorbed to produce electronically excited states or remove an electron, the absorption intensity being the measurable quantity. 
\begin{table}[h]
    \begin{center}
\begin{tabular}{| c | c | c | c | c |}
\hline
 molecule & initial state & final state & spectroscopy type & main/SI \\ 
 \hline
 formic acid & neutral, singlet & cation, doublet & photoelectronic & main \\  
 p-benzyne & anion, doublet, & neutral, triplet & NIPES & main \\
 p-benzyne & anion, doublet & neutral, singlet & NIPES & SI \\
 methylene & anion, doublet & neutral, triplet & NIPES & SI \\
 methylene & anion, doublet & neutral, singlet & NIPES & SI \\
 pyridazine & ground state, singlet & first excited state, singlet & UV-vis & main \\
 formaldehyde & ground state, singlet & first excited state, singlet & UV-vis & main \\
 PNA & ground state, singlet & first excited state, singlet & UV-vis & SI \\
 naphthalene & ground state, singlet & first excited state, singlet & UV-vis & SI \\
 \hline
\end{tabular}
\caption{Overview of the molecules and their transitions investigated in this work.}
\label{table:overview}
\end{center}
\end{table}

The data (molecular parameters and computed Franck-Condon profiles) will be made available upon publication. 
For formic acid we have used previously reported data \cite{huh2015boson}, based on the raw data from reference \cite{jankowiak2007vibronic}, since our calculations perfectly reproduce these results (see below). In our measurements, we have used seven modes to aid the comparison to previous work \cite{huh2015boson}. Molecular structures and inputs are available from the authors upon reasonable request.

More technical details on quantum chemical calculations of the spectra are given below.

    \subsection{Photoelectron spectroscopy}
    In photoelectron spectroscopy a photon is used to remove an electron from the neutral atom. This implies a vibronic transition from the lowest ground state of the neutral molecule to the ground state of the cation. To test this transition we have selected formic acid as it has been extensively studied in previous photonic investigations. To aid in comparing our results to prior work, we have selected the same theoretical methodology for ab initio calculations as in the article from which the data for photonic computing has been derived \cite{jankowiak2007vibronic}: the structure and harmonic vibrations of formic acid and its cation were optimized using CCSD(T) \cite{raghavachari1989fifth} with the cc-pVTZ basis set \cite{dunning1989gaussian} using the ORCA software package \cite{ORCA5}. Since analytical Hessian for CCSD(T) is not available it was calculated numerically. Our data are identical to those in the reference up to negligible numerical deviations.

    \subsection{Negative ion photoelectron spectroscopy (NIPES)}
    NIPES is similar to photoelectron spectroscopy, the difference being that an electron is removed from a negatively charged ion. This approach is typically used to investigate different electronic states belonging to different multiplicities of neutral molecule, as depending upon which valence electron is removed, different electronic states can form \cite{wenthold1999negative}. Two famous diradical molecules that were experimentally investigated using this method are p-benzyne and methylene. These two molecules have close lying singlet and triplet states. NIPES provides a straightforward approach to not only measure the energy difference but also investigate the vibronic transitions. As these molecules have semi-degenerate orbitals it is recommended to investigate the system with a method capturing static electron correlation. Therefore the structure optimization and calculation of harmonic vibrations were done with the complete active space self-consistent field (CASSCF) method \cite{SUN2017291}, with def2-TZVP (p-benzyne) and def2-TZVPD  \cite{B515623H} basis sets (with the corresponding resolution-of-identity auxiliary basis set) using PySCF software package \cite{sun2007python}. The active space sizes were 8 electrons in 8 orbitals for p-benzyne; 6 electrons in 6 orbitals for methylene.

    \subsection{Ultraviolet–visible spectroscopy (UV-vis)}
    In UV-vis spectroscopy a beam of light is passed through the sample. Photons absorbed by the sample promote electrons from valence orbitals and create excited states. The amount of light absorbed at each wavelength is measured to produce the corresponding absorption spectrum. We have selected formaldehyde, naphthalene, para-nitroaniline (PNA), and pyridizine molecules for frequency calculations. The calculations are done using time-dependent density functional theory with the hybrid B3LYP \cite{becke1993new} functional using def2-TZVP \cite{B515623H} basis set.

    \section{Benchmarking GBS against classical simulation}
    To ensure the validity of the parameters obtained for the GBS simulation, we compare the GBS simulations to classical simulations performed with FCClasses \cite{FCClasses}.
    Since we do not investigate the validity of the GBS approach in general, and the energies for the classical simulation have slight offsets due to numerical approximations, we do not calculate the similarities and instead show only qualitative agreement.
    
\begin{figure}
        \centering
        \includegraphics[width=\textwidth]{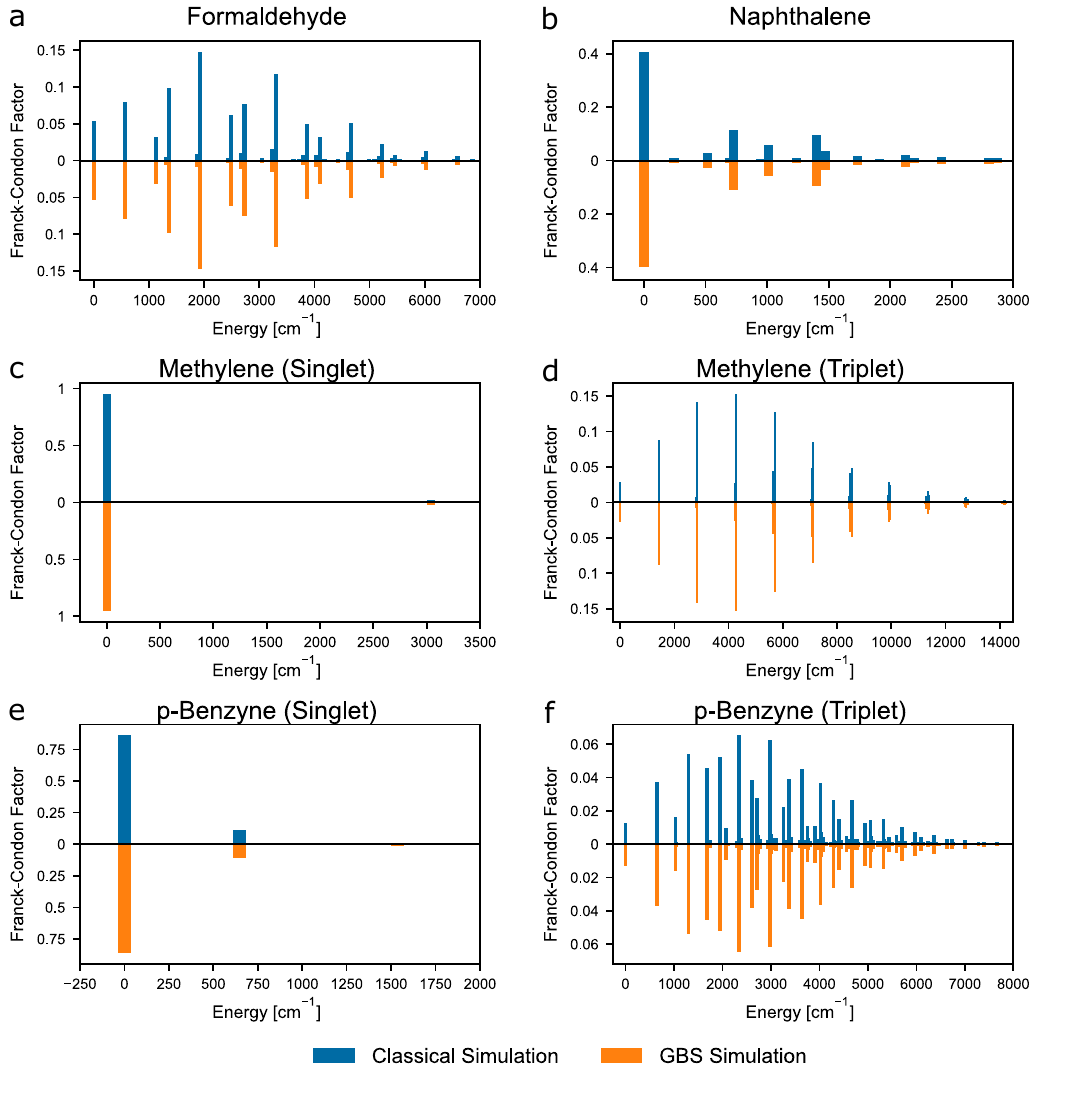}
        \caption{Comparison between classical simulations with FCClasses (blue) and GBS simulations (orange) for various molecules.}
        \label{fig:classical1}
    \end{figure}

    The comparison between the classical simulations with FCClasses and the GBS simulation for simpler molecules (formaldehyde, naphthalene, methylene and p-benzyne) can be seen in Fig.~\ref{fig:classical1}.
    Classical and GBS simulations both lead to vibronic spectra with the same overall structure.
    Slight differences in the FCF values can be observed in some cases, for example the $0$ energy peak in naphthalene is slightly higher for the classical simulation. However, these differences are small and can be attributed to the classical simulations neglecting some of the less likely transitions. Overall, the comparison validates our GBS simulation.

\begin{figure}
        %\ASF benchmarkcentering
        \includegraphics[width=\textwidth]{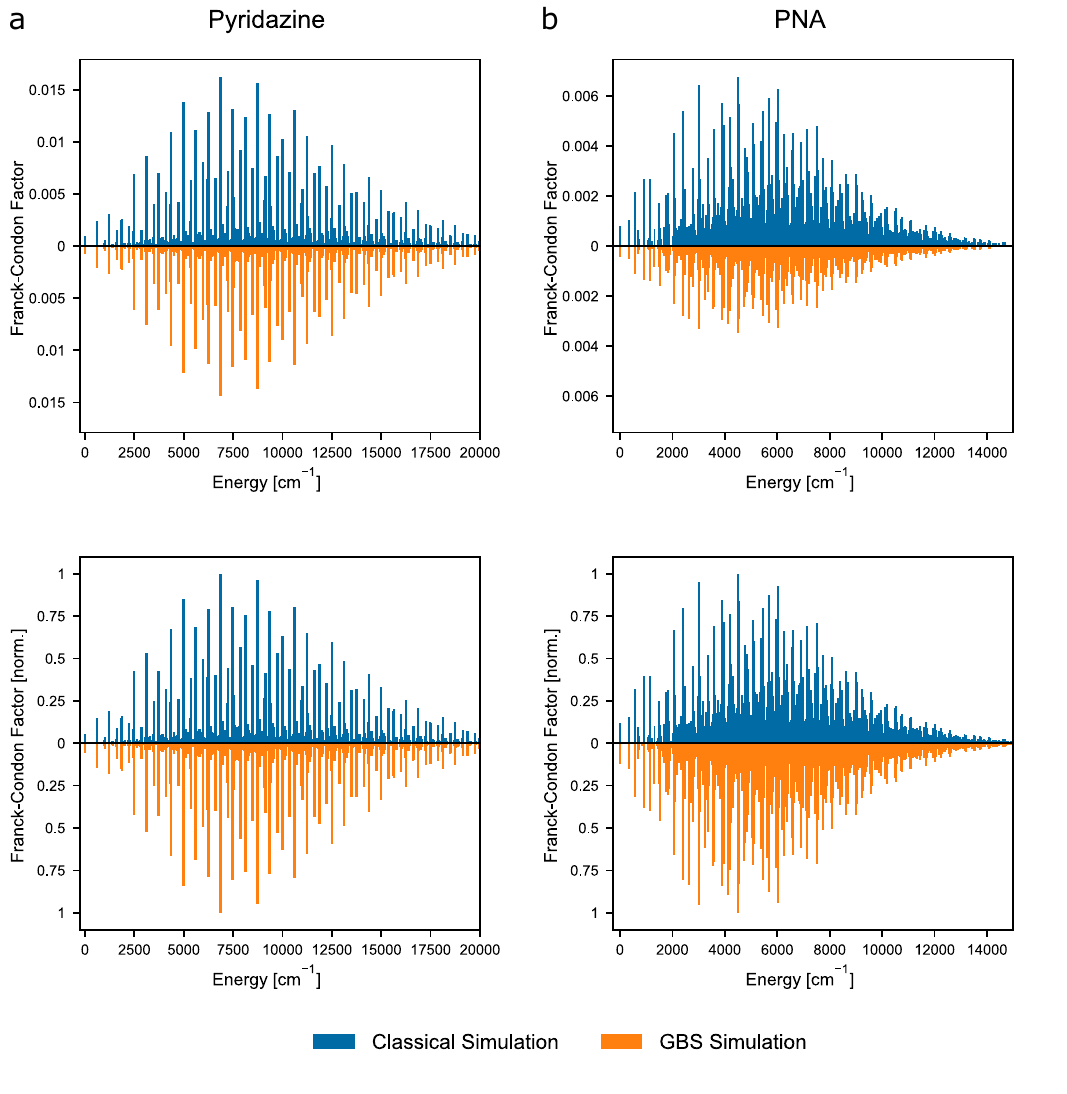}
        \caption{Comparison between classical simulations with FCClasses (blue) and GBS simulations (orange) for pyridazine (a) and PNA (b).
        The bottom row shows the same spectra but with FCFs normalized to $1$ for a better comparison of the structure.}
        \label{fig:classical2}
    \end{figure}

    For pyridazine and PNA this comparison becomes more complicated due to the complexity of these two molecules. In the case of pyridazine (Fig.~\ref{fig:classical2} a) the classical simulation leads to slightly higher values than the GBS simulation. However, when normalizing both spectra one can see that the overall shape of the spectrum between both simulations matches well, as shown in bottom panel of Fig.~\ref{fig:classical2}. Similar to the previous case this can be attributed to the classical simulation neglecting less likely transitions which now, due to the larger complexity of the molecule, has a more significant effect on the spectrum.
    For PNA (Fig.~\ref{fig:classical2} b) we can observe that this is even more pronounced, with the FCFs differing by  a factor of approximately $2$.

    Note that formic acid has not been explicitly shown here since the validity of those parameters was already shown by Huh \textit{et al} \cite{huh2015boson}. 
    
    \section{Detailed experimental setup}
\begin{figure}[h!]
        \centering
        \includegraphics[width=0.9\textwidth]{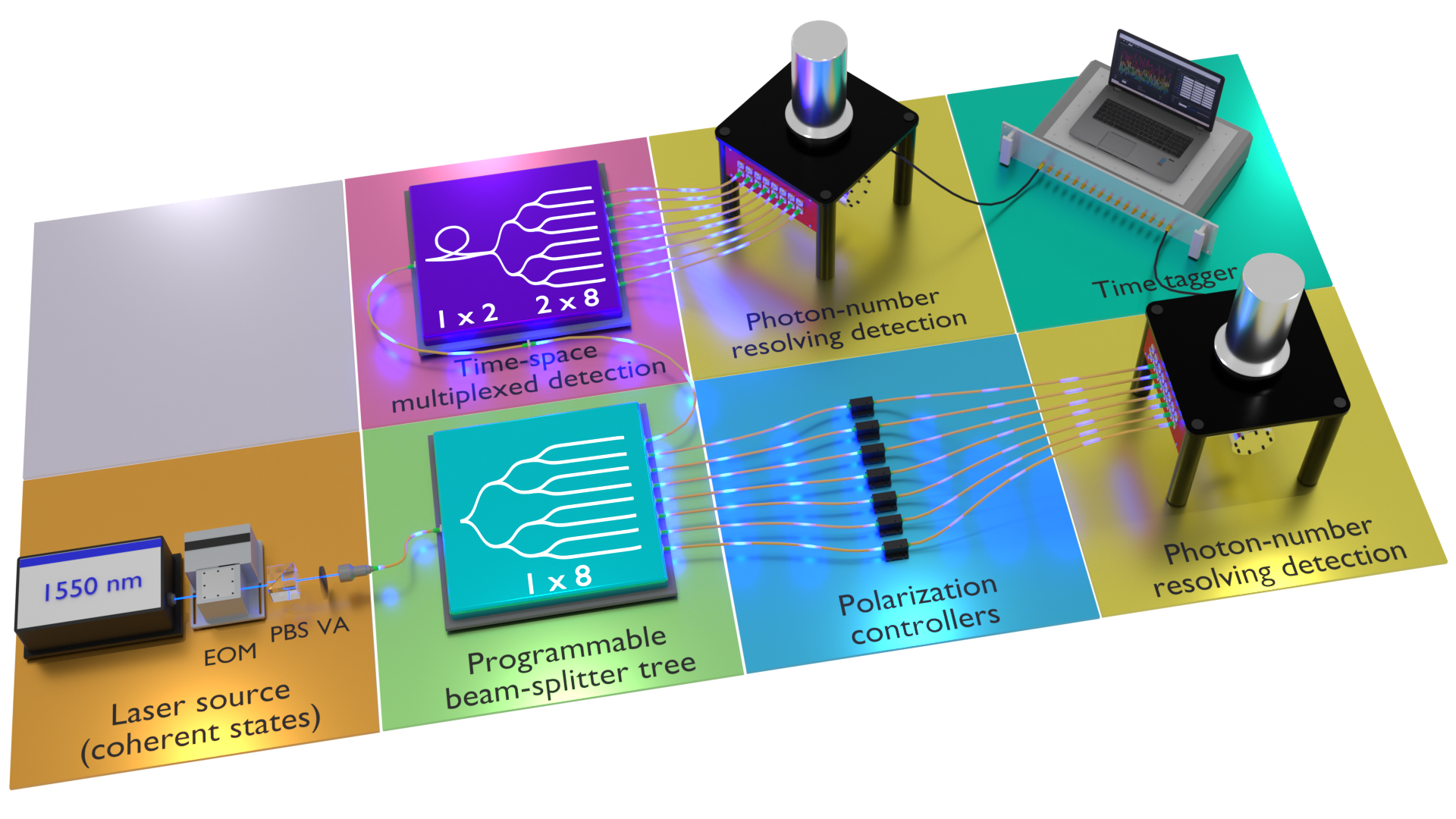}
        \caption{Detailed schematic of the experimental setup for sampling vibronic spectra using the linear coupling approximation. See supporting text for details. EOM: electro-optic modulator, PBS: polarizing beam-splitter, VA: variable attenuator.}
        \label{fig:FullSetup}
    \end{figure}

    The following is a detailed description on our experimental setup for simulating vibronic spectra under the linear coupling approximation (see Fig.~\ref{fig:FullSetup}). We used an Er-doped fiber mode-locked laser frequency comb (Menlo Systems GmbH, SmartComb) operating at 1550/,nm central wavelength as a source of coherent states. The laser has 80 MHz repetition rate, which was reduced to 1 MHz by a combination of electro-optic modulator (QUBIG GmbH, HVOS-SWIR) and polarizing beam-splitter. A free-space variable attenuator was employed to adjust the overall mean photon number. The attenuated laser beam was then coupled to an optical fiber and directed to a programmable beam-splitter tree (QuiX Quantum, photonic processor). This device allows us to split the beam into multiple outputs with splitting ratios that could be individually controlled.
    
    The output modes with mean photon number $|\beta_i|^2\leq0.35$ were connected to fiber coupled polarization controllers. The polarization controllers were used to fine-tune the mean photon number by rotating the polarization of the field reaching the polarization sensitive superconducting nano-wire single-photon detectors (SNSPDs, Single Quantum). A single mode with $|\beta_i|^2\geq0.35$ could be connected to the time-space multiplexed detection unit, allowing for detection of states with signficant higher photon number contributions. This unit is a passive beam-splitter tree that redistributes a relatively large mean photon number into 2 time- and 8 spatial-bins. This configuration employs 8 additional SNSPDs (spatial bins) that could be used twice (time bins) for each sample. 
    In combination with intrinsic photon number resolution \cite{IntrinsicPNR}, this technique allows for resolving up to $40$ photons in one mode. 
    The reason that it does not reach $48$ as one would expect for an intrinsic PNR of $3$ is that for some bins the intrinsic PNR can only resolve up to $2$ photons. 
    This is caused by a slight polarization walk-off due to imperfect fiber splicing and a limitation in the PNR calibration capabilities.
    Finally, each SNSPD's output was connected to a time tagger (Swabian Instruments GmbH). The time tagger data were analyzed using a program that generated the final statistics to calculate the spectra.

    \section{Additional results}
    In addition to the results presented in the main paper, simulations and measurements for other molecules/transitions were taken.
    As described in the main paper, each measurement run consists of approximately $18 \times 10^7$ samples.
    With the same conditions, we measured only modes with displacements $|\beta_i|^2\geq10^{-4}$. Modes with $|\beta_i|^2\geq0.35$ were measured using the time-space multiplexed detection for additional photon-number resolution. Note that for some molecules, i.e. where there were multiple modes with larger displacement, the measurements were divided into multiple runs:
    \begin{itemize}
        \item p-benzyne (singlet): 1 measurement with 3 modes
        \item methylene (singlet): 1 measurement with 2 modes 
        \item methyle (triplet): 2 measurements with 1 mode each
        \item naphthalene: 1 measurement with 8 modes
        \item PNA: 4 measurements with 6, 5, 5 and 4 modes, respectively
    \end{itemize}
    This splitting can be done since, as described in the manuscript, the modes 
    in the linear coupling approximation are independent.
    
    Simulations of molecules with multiple measurement runs were analyzed separately. The energies of the separate runs were summed and the corresponding probabilities were multiplied in all possible combinations to reconstruct the spectrum.
    It is important to note that when combining the $4$ measurement runs of PNA, we discarded all probabilities $<10^{-6}$ to reduce the amount of data points and speed up computation time. This leads to $\approx0.01$~\% of the spectrum being discarded. 
    
    Simulations for the parallel approximation and GBS were also performed, generating $18\times 10^7$ and $2.5\times 10^6$ samples, respectively.
    When calculating the spectra, two energies $E_i$ and $E_j$ with $|E_i-E_j| \leq 10^{-10}$ are assumed to be equal and binned to the same energy. This only affects the spectra of PNA.
    The same condition for equal energies is used in the calculation of the similarities. 
    
\begin{figure}[h!]
        \centering
        \includegraphics[width=\textwidth]{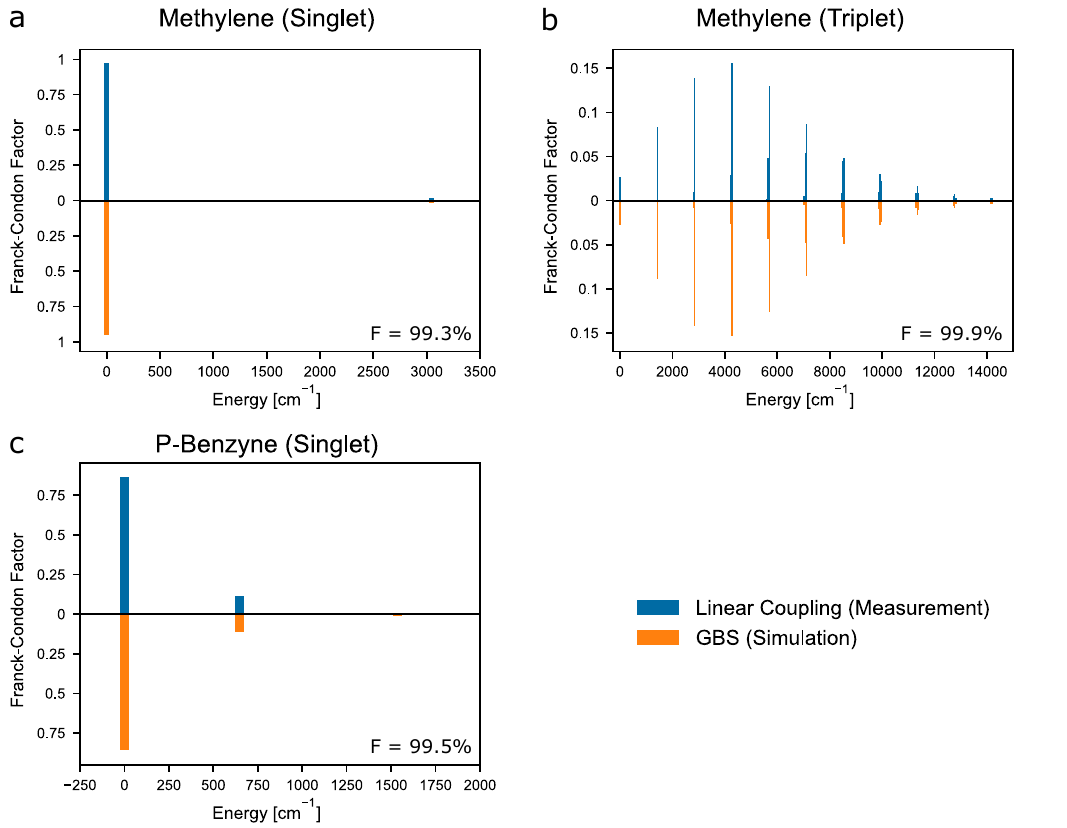}
        \caption{Comparison between spectra determined using the linear coupling approximation measured in our photonic system (blue) and GBS simulations (orange). All of these spectra can be accurately reconstructed using the linear coupling approximation. See text for details.}
        \label{fig:AR1}
    \end{figure}

    The results for the singlet transitions of methylene (Fig.~\ref{fig:AR1} a) and p-benzyne (Fig.~\ref{fig:AR1} c) show relatively simple spectra with few peaks. We can see that for both the linear coupling approximations performs well, reaching similarities of $99.3$~\% and $99.5$~\%, respectively.
    For the triplet transition of methylene (Fig.~\ref{fig:AR1} b) the spectrum shows more features. In this case the photonic measurement, i.e. linear coupling, also leads to almost perfect results with a similarity of $99.9$~\%.

\begin{figure}[h!]
        \centering
        \includegraphics[width=\textwidth]{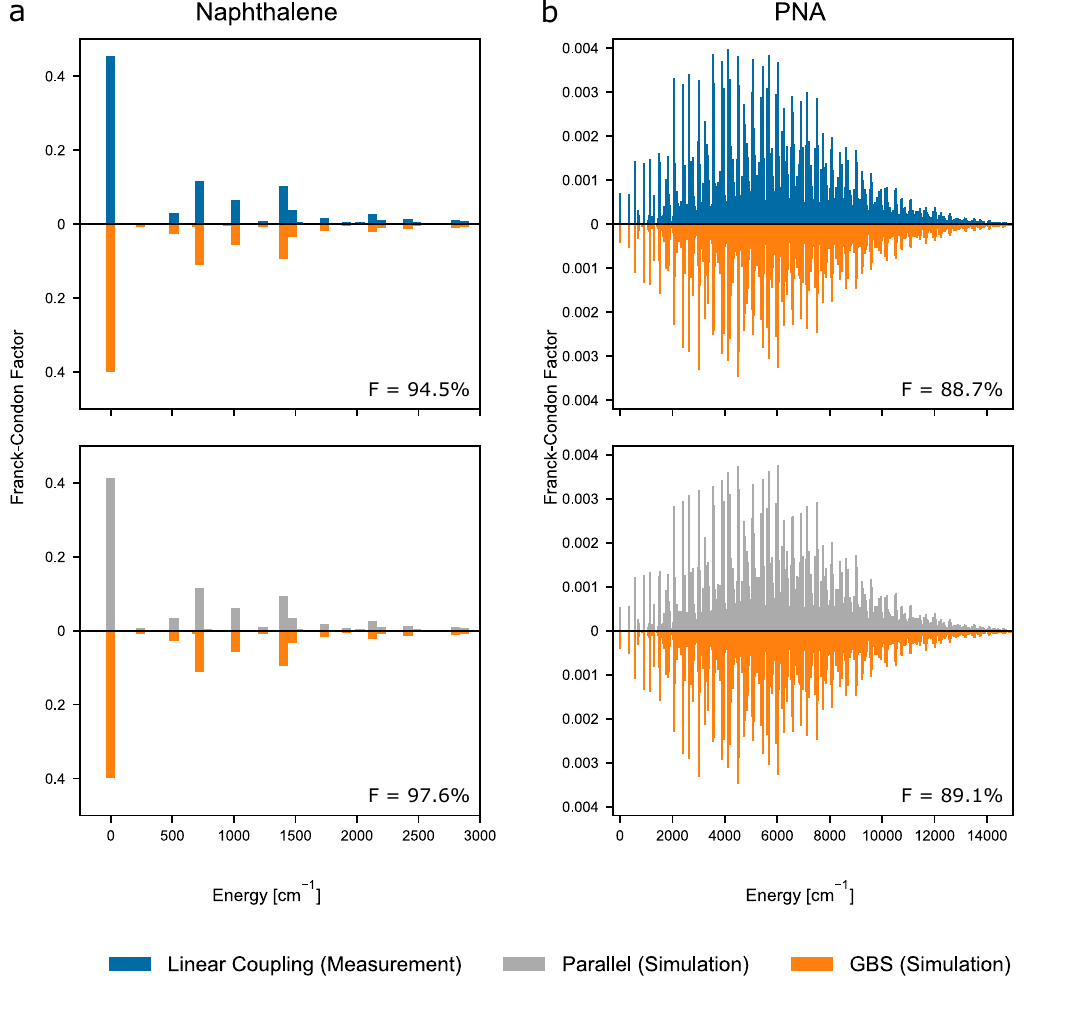}
        \caption{Comparison between spectra in the linear coupling approximation measured in our photonic system (blue), simulated parallel approximation (gray) and GBS simulation (orange) for naphthalene (a) and PNA (b).  
        See text for details.}
        \label{fig:AR2}
    \end{figure}

    Fig.~\ref{fig:AR2} shows the results for naphthalene and PNA. For napththalene (Fig.~\ref{fig:AR2} a) linear coupling and parallel approximations both lead to decent similarities of $94.5$~\% and $97.6$~\%, respectively. A GBS approach would only lead to a small improvement, primarily noticeable in the FCF at $0$ energy. 
    PNA is an example of a molecule where neither linear coupling nor parallel approximation lead to accurate results, with $88.7$~\% and $89.1$~\% similarity, respectively. While for the example of pyridazine in the main paper the inaccuracy was mostly noticeable in terms of the actual values of the FCFs, here significant changes in the shape of the spectrum can also be observed.

    \section{Effect of imperfect displacements and neglected modes}
    To investigate the accuracy of our setup, we quantified how accurately we were able to set the displacements.
    From our measurement we determined the measured mean photon number $\langle n_i\rangle_{Meas.}$ for all implemented vibronic modes and calculate the deviation from the target mean photon numbers $\langle n_i \rangle = |\beta_i|^2$.
    Next, we calculated the percentage error
    \begin{equation}
        \Delta n_i = \frac{|\langle n_i\rangle_{Meas.}-\langle n_i \rangle|}{\langle n_i \rangle},
    \end{equation}
    from which the mean error weighted by the mean photon numbers was calculated
    \begin{equation}
        \Delta n = \frac{\sum_i \langle n_i\rangle\cdot\Delta n_i}{\sum_i \langle n_i\rangle}.
    \end{equation}
    The resulting weighted mean errors are collected in the second column of table~\ref{table:error}.
    
    To investigate the effect of neglecting modes with $|\beta_i|^2 < 10^{-4}$ and imperfect values for $\beta_i$ we performed additional simulations. For that purpose we (numerically) generated $18\cdot10^7$ samples for the linear coupling approximation, using the perfect displacement values and considering all vibronic modes of the molecules.
    We then calculated the similarity between the simulated linear coupling and our photonic measurement to quantify how close our measurement is to a more ideal case. Additionally, we calculated the similarities between the simulated linear coupling and GBS simulation to investigate the effect of our experimental uncertainties on the similarity compared to GBS.
    \begin{table} [h!]
    \begin{center}
    \scalebox{0.9}{
    \begin{tabular}{| c | c | c | c | c |}
    \hline
     molecule & $\Delta n$ [\%] & Sim. LC-Meas.[\%, $\pm0.01$] & Meas.-GBS [\%, $\pm0.04$] & Sim. LC-GBS [\%, $\pm0.04$]\\ 
     \hline
     \textcolor{violet}{formic acid} & $0.53$ & $>99.99$ & $98.39$ & $98.39$\\ 
     \textcolor{violet}{p-benzyne (S)} & $0.60$ & $>99.99$ & $99.54$ & $99.54$\\
     \textcolor{violet}{p-benzyne (T)} & $0.76$ & $99.99$ & $99.53$ & $99.54$\\
     \textcolor{violet}{methylene (S)} & $1.60$ & $>99.99$ & $99.29$ & $99.26$\\
     \textcolor{violet}{methylene (T)} & $0.85$ & $99.99$ & $99.88$ & $99.89$\\
     \textcolor{orange}{pyridazine} & $1.75$ & $99.96$ & $75.92$ & $76.02$\\
     \textcolor{cyan}{formaldehyde} & $4.46$ & $99.94$ & $96.53$ & $96.51$\\
     \textcolor{orange}{PNA} & $0.58$ & $99.81$ & $88.74$ & $88.69$ \\
     naphthalene & $0.63$ & $99.99$ & $94.45$ & $94.47$\\
     \hline
    \end{tabular}
    }
    \caption{Overview of weighted mean photon number error ($\Delta n$) and similarities for the investigated molecules. (S) and (T) indicate the singlet and triplet transitions, respectively.
    Sim. LC: Simulated linear coupling approximation, Meas.: Measured linear coupling approximation, GBS: Gaussian boson sampling simulation. The spectra of the molecules marked in violet are well described by linear coupling, the one in cyan by parallel and the ones in orange by full Duschinsky/GBS. Naphthalene is kept in black since its shape is well described by linear coupling and parallel already, but the similarity is relatively low.}
    \label{table:error}
    \end{center}
    \end{table}
    
    Table~\ref{table:error} shows that our photonic measurement agrees almost perfectly with the simulated linear coupling approximation.
    For some smaller molecules (formic acid, p-benzyne (singlet) and methylene (singlet)) similarities of $>99.99$~\% can be observed. 
    The worst similarity can be seen for the most complex molecule, PNA, showing a similarity of $99.81$~\%.
    When comparing both linear coupling approaches to the GBS simulation it can be seen that they match well within the given uncertainty of $0.04$~\%. Therefore, we can deduce that our experimental uncertainties and approximations have a negligible effect on the results.

    \section{Duschinsky matrix for formic acid}
\begin{figure}[h!]
        \centering
        \includegraphics[]{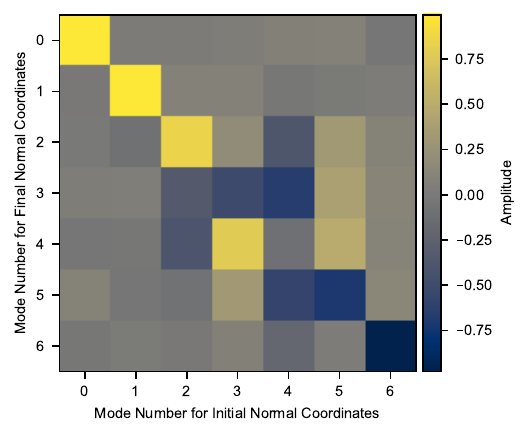}
        \caption{Duschinsky matrix for formic acid considering 7 vibrational modes.}
        \label{fig:form_dusch}
    \end{figure}

    Fig.~\ref{fig:form_dusch} shows the Duschinsky matrix $\mathbf{U}$ for the $7$ considered vibronic modes of formic acid.
    As discussed in the main paper, mode mixing in some of the modes can be observed. 
    However, the squeezing parameters arising from the Duschinsky matrix and frequency change is relatively small, leading to only small changes in the vibronic spectrum as discussed.
    \newpage
    \bibliographystyle{naturemag.bst}
	\bibliography{References2.bib}